\documentclass[%
 reprint,
 amsmath,amssymb,
 aps,
prstab,
showkeys,
superscriptaddress
]{revtex4-2}

\usepackage[normalem]{ulem}
\usepackage{hyperref}
\usepackage{gensymb}
\hypersetup{
    colorlinks=true,
    linkcolor=blue,
    filecolor=magenta,      
    urlcolor=blue,
    citecolor = blue,
    bookmarks=false,
}
\usepackage{hhline}
\usepackage{graphicx}
\usepackage{dcolumn}
\usepackage{bm}

\usepackage[mathlines]{lineno}

\newcommand{\subsc}[1]{\ensuremath{_{\textrm{\begin{scriptsize}#1\end{scriptsize}}}}}
\newcommand{\supsc}[1]{\ensuremath{^{\textrm{\begin{scriptsize}#1\end{scriptsize}}}}}

\begin{document}

\title {Mid-T Heat Treatments on BCP'ed Coaxial Cavities at TRIUMF} 

\author{P. Kolb}
 \email[Email: ]{kolb@triumf.ca}
 \affiliation{TRIUMF, 4004 Wesbrook Mall, Vancouver, B.C. V6T 2A3, Canada}
\author{Z. Yao}
 \affiliation{TRIUMF, 4004 Wesbrook Mall, Vancouver, B.C. V6T 2A3, Canada}
\author{A. Blackburn}
 \affiliation{University of Victoria, Victoria, B.C., Canada}
 \author{R. Gregory}
 \affiliation{University of Victoria, Victoria, B.C., Canada}
 \affiliation{TRIUMF, 4004 Wesbrook Mall, Vancouver, B.C. V6T 2A3, Canada}
\author{D. Hedji}
 \affiliation{University of Victoria, Victoria, B.C., Canada}
 \affiliation{TRIUMF, 4004 Wesbrook Mall, Vancouver, B.C. V6T 2A3, Canada}
 \author{M. McMullin}
 \affiliation{University of Victoria, Victoria, B.C., Canada}
 \affiliation{TRIUMF, 4004 Wesbrook Mall, Vancouver, B.C. V6T 2A3, Canada}
 \author{T. Junginger}
 \affiliation{University of Victoria, Victoria, B.C., Canada}
 \affiliation{TRIUMF, 4004 Wesbrook Mall, Vancouver, B.C. V6T 2A3, Canada}
 \author{R.E. Laxdal}
 \affiliation{TRIUMF, 4004 Wesbrook Mall, Vancouver, B.C. V6T 2A3, Canada}
 
\date{\today}	

\begin{abstract}
Mid-T heat treatments in the range from 250 to 400\,$^\circ$C on superconducting radio-frequency (SRF) cavities have been shown to provide high quality factors that rise with applied rf field strength in high frequency, electro-polished (EP), elliptical cavities operating at 2\,K, similar to nitrogen doped cavities. The rise in quality factor is attributed to a decrease in the temperature dependent part of the surface resistance R\subsc{BCS}. Until now, no results have been reported for these new treatments on quarter-wave resonators (QWR) and half-wave resonators (HWR). The TRIUMF multi-mode coaxial cavities are dedicated test cavities that allow frequency and temperature resolved performance characterization of treatments without changing environments, therefore providing an excellent test vehicle to test these new treatments with rf frequencies ranging from 200 to 1200\,MHz.

In this paper, performance measurements from both QWR and HWR cavities are reported and their performance compared with four different treatments: baseline, a conventional 120\,$^\circ$C low temperature bake for 48 hours, and two mid-T bakes at 300 and 400\,$^\circ$C for 3 hours. In addition, sample analysis using SEM, EDX and SIMS of witness samples is also shown.

It is found that the mid-T bakes are not directly transferable to low frequency cavities. In the fundamental modes of the two test cavities, no performance gain over the baseline treatment nor a decreasing temperature dependent component with rising rf amplitude was observed.  At frequencies above 1\,GHz and low temperatures, the mid-T bakes show a reduced field dependence of R\subsc{BCS} compared to both the baseline and 120\,$^\circ$C treatments. 
\end{abstract}
\maketitle
\section{Introduction}

Modern nuclear physics experiments, such as the experiments conducted at the ISAC-II \cite{ISAC} facility at TRIUMF, often rely on heavy ion particle accelerators for high energy beams. To deliver these beams in an efficient way often superconducting radio-frequency (SRF) technology is used. A major cost driver for SRF based accelerators are cryogenic losses in SRF cavity walls caused by the rf field used to accelerate the beam. The required cryogenic infrastructure and liquid helium (LHe) as coolant are costly on capital and operational budgets. Cryogenic losses are inversely proportional to the unloaded quality factor Q\subsc{0}. Therefore, increasing the quality factor in SRF cavities is essential to reduce the cryogenic load and therefore reduce operating costs or design accelerators with smaller cryoplants. This is most relevant for SRF based accelerators with a high duty factor, where the active load dominates over the static load. Additionally, increasing the operational accelerating fields, and therefore usable voltage, can lead to shorter, more compact accelerators, reducing real estate expenses.

Several novel heat treatments that increase Q\subsc{0} more than conventional treatments have been been developed and applied to 1.3\,GHz single cell elliptical cavities \cite{Grassellino:Infusion, Grassellino:Doping, gonnella:srf2019-mop045}, all on electro-polished (EP) cavities. Coaxial style cavities such as quarter-wave and half-wave resonators (QWR and HWR respectively) used in heavy ion accelerators are often treated with buffered chemical polishing (BCP) due to the more complex geometry and the resulting technical challenges of using EP.

\begin{figure}
\centering
\includegraphics[width=0.95\linewidth]{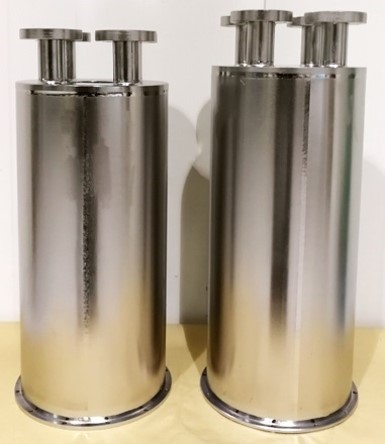}
\caption{\label{Fig_cavities}Photo of the HWR (left) and QWR (right) cavities used in this study. }
\end{figure}

Recently discovered heat treatments at 250 to 500\,$^\circ$C in elliptical 1.3\,GHz cavities with an EP surface finish have been shown to produce comparable performance to nitrogen doped cavities \cite{He_2021,10.1093/ptep/ptab056, steder:linac2022-thpoge22}, the current state-of-the-art treatment to date for high Q\subsc{0} at moderate gradients \cite{DHAKAL2020100034}, while being a simpler cavity processing recipe. The performance increase is a result of oxygen dissolution and diffusion of the native niobium oxide layers \cite{10.1063/5.0059464}. At temperatures between 250 and 500\,$^\circ$C the Nb\subsc{2}O\subsc{5} layer dissolves and regrows back shallower after air exposure \cite{PhysRevAccelBeams.22.103102}. This avoids using an external doping agent like nitrogen. 
\begin{figure}[t]
    \centering
    \includegraphics[width=0.95\linewidth]{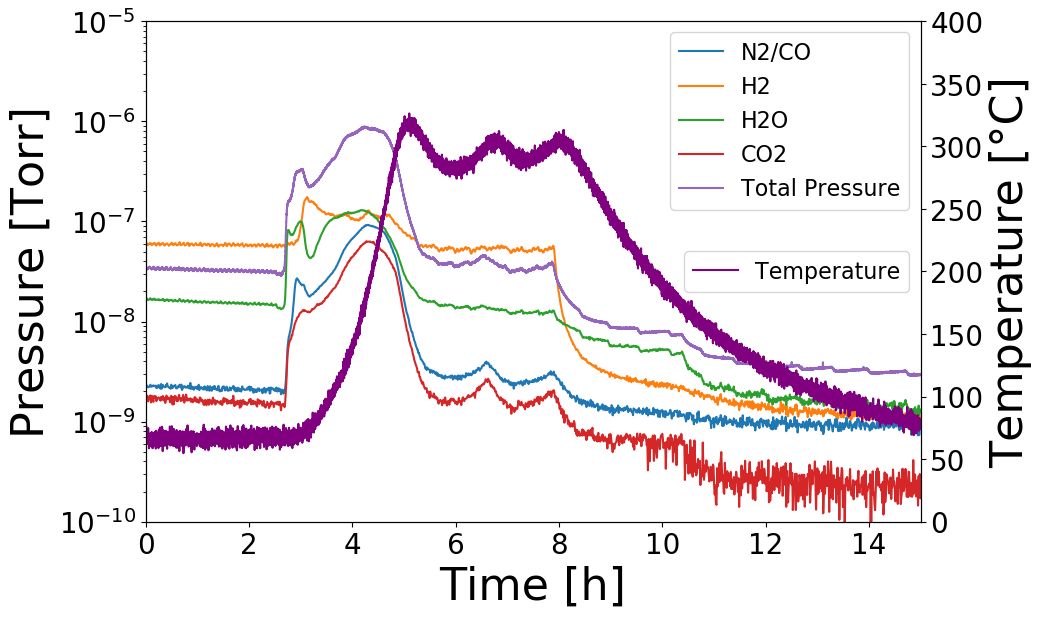}
    \caption{\label{Fig_QWR300C}Temperature and residual gas profiles during the heat treatment of the QWR at 300\,$^\circ$C. The other mid-T treatments follow similar patterns. The fluctuations around goal temperature are due to the slow response of the cavity temperature regulation.}
\end{figure}

These mid-T or 'oxygen diffusion' treatments allow quicker processing compared to N\subsc{2}-doping as they eliminate the injection of N\subsc{2} into the furnace at high temperatures and a post-doping EP step. This simplified process decreases processing time over N\subsc{2}-doping and should also increase consistency in larger cavity productions. Since mid-T bakes show a decreasing R\subsc{BCS} with increasing rf field amplitude, this treatment is of high interest for low frequency SRF cavities that typically operate at a temperature of 4.2\,K. At this operating temperature, the surface resistance R\subsc{s} is dominated by its temperature dependent component R\subsc{BCS} and the temperature independent part R\subsc{Res} contributes only a small fraction. A decreasing R\subsc{BCS} at 4\,K would be an enormous benefit for SRF based heavy ion accelerators operating in continuous wave (cw) mode at moderate peak surface fields, such as the ISAC-II or FRIB accelerators. The first reported results on BCP'ed 1.3\,GHz nine-cell cavities \cite{Yang_2023} do not show the characteristic increase in Q\subsc{0}, but still a larger Q\subsc{0} than a baseline treatment. This is similar to the difference between EP'ed and BCP'ed N\subsc{2}-doped cavities \cite{PhysRevAccelBeams.22.122002}.
Similarly, results on EP'ed 650\,MHz ellitpical cavities with either N\subsc{2}-doping \cite{10.1063/5.0068531} or a 300\,$^\circ$C bake \cite{app12020546} do not show the increasing Q\subsc{0}, suggesting that there is a frequency dependence to both treatments. 

In this paper, the effects of two mid-T bakes at 300 and 400\,$^\circ$C on a QWR and a HWR, both with BCP surface finishes, are studied and presented. The cavities are shown in Fig. \ref{Fig_cavities}.

This paper presents the used methodology in Sec. \ref{secMethodology}, results of cavity measurements and sample studies in Sec. \ref{secResults} and findings are discussed in Sec. \ref{secDiscussion} before summarizing them in Sec. \ref{secSummary}.

\section{Methodology\label{secMethodology}}
Two dedicated research cavities are used in this study: one QWR and one HWR, both shown in Fig. \ref{Fig_cavities}. The two cavities and many aspects of the data acquisition and processing are described in detail in \cite{PhysRevAccelBeams.23.122001}. These cavities are simplified versions of the QWR and HWR geometries and designed for excitation of not only the fundamental rf mode, but also higher order TEM modes. This allows the extraction of the frequency dependence of key performance parameters such as quench field and Q\subsc{0}, defined as
\begin{align}
Q_0 = \frac{\omega U}{P}
\end{align}
with $\omega$ as resonant frequency, $U$ as stored energy in the RF field, and $P$ as dissipated power in the cavity walls, without changing to different cavities or different testing parameters such as cooldown speed or external magnetic fields. These two cavities cover a frequency range from 220\,MHz to 1.2\,GHz, covering commonly used frequencies for SRF cavities.
\begin{figure}[th]
	\centering
    \includegraphics[width=0.95\linewidth]{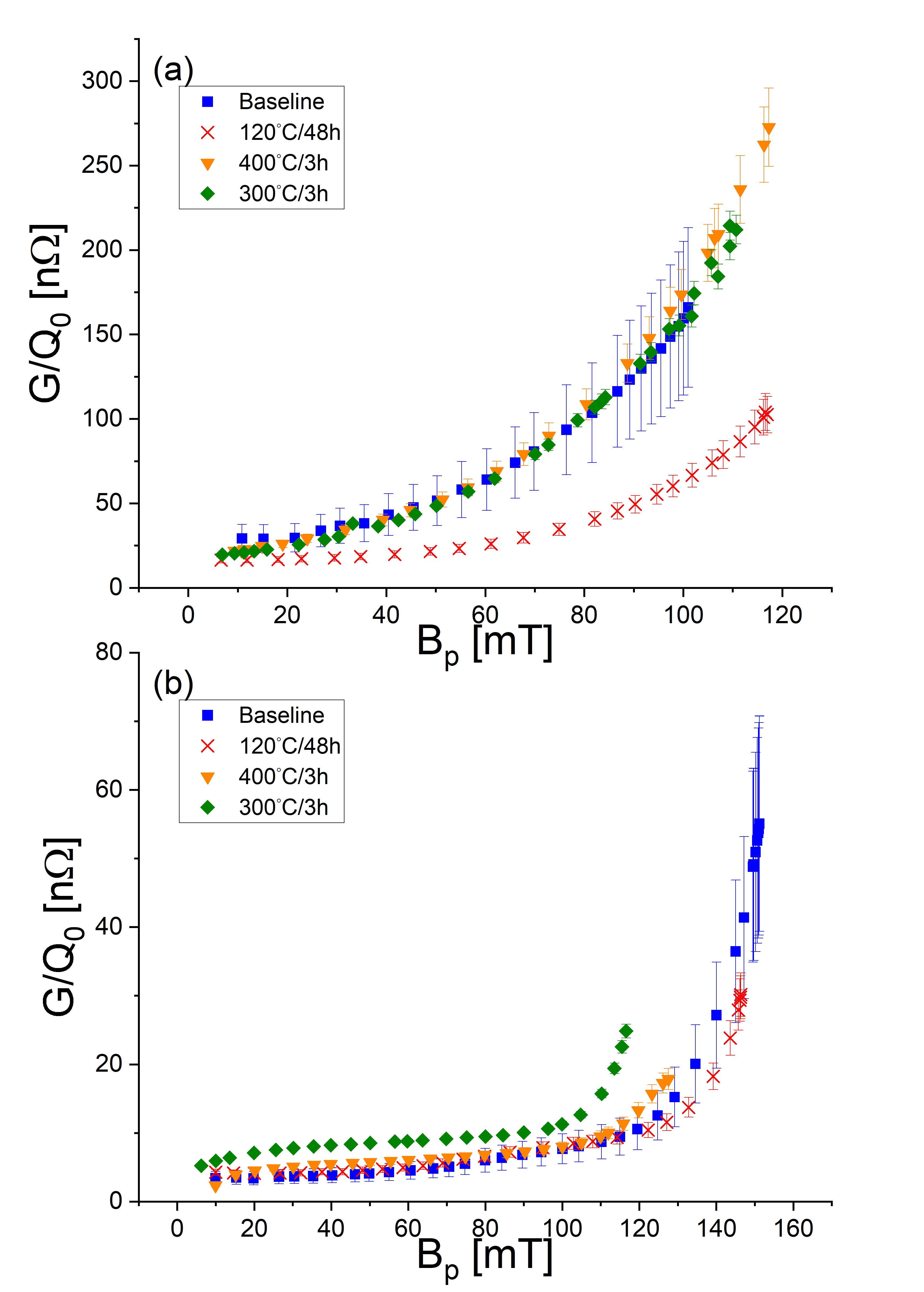}
	\caption{\label{Fig_QWR217MHz_Q_Curve}(a) At 4\,K the 220\,MHz mode of the QWR  with the 120\,$^\circ$C bake performs significantly better with a lower surface resistance over a large range field amplitudes up to the quench field. (b) At 2\,K the mid-T bakes increase G/Q\subsc{0} compared to the baseline and 120\,$^\circ$C bake.}
\end{figure}

This work compares performance of the two test cavities with a baseline treatment (bulk BCP - 800\,$^\circ$C degassing - flash BCP), a commonly used low temperature bake at 120\,$^\circ$C for 48\,hours, and two mid temperature bakes at 300 and 400\,$^\circ$C for 3\,hours. Between the 120\,$^\circ$C, 400\,$^\circ$C, and 300\,$^\circ$C heat treatments the surface is reset with a 20\,$\mu$m etch via BCP. Measurements of Q\subsc{0} as a function of peak magnetic surface field B\subsc{P} are done at 4.2\,K and 2.0\,K, as well as measurements of Q\subsc{0} during the cooldown between the two temperatures at selected field amplitudes. This allows for the deconvolution of the surface resistance R\subsc{s} into its temperature dependent and independent components, R\subsc{BCS} and R\subsc{Res} respectively, by fitting the data to
\begin{align}
R_s(T) &= \frac{a_0}{T}\ln\left(\frac{4k_bT}{\hbar\omega}\right)\exp\left(-\frac{a_1(T)T_c}{T}\right) + a_2 \label{Eq_Rs}\\
&= R_{BCS}(T) + R_{Res}
\end{align}
which is a parameterized version of the BCS surface resistance \cite{Gurevich_2017} with $a_0$, $a_1$ and $a_2$ as fit parameters. The energy gap $\Delta$ is represented by $a_1$ and its temperature dependence is modelled \cite{HalbritterFortran} as
\begin{align}
\frac{a_1(T)}{a_1(0\,K)} = \sqrt{\cos\left(\frac{\pi}{2}\left(\frac{T}{T_c}\right)^2\right)}.
\end{align}

Further details of the test setup and data processing, including accounting for the non-uniform field distribution across the surface \cite{PhysRevAccelBeams.21.122001}, are described in \cite{PhysRevAccelBeams.23.122001}. Measurements of Q\subsc{0} and B\subsc{p} are done in a zeroed external magnetic field using a set of Helmholtz coils which surround the cavity and cryostat. In this paper, the performance is often presented as G/Q\subsc{0} with G as the geometry factor defined as
\begin{align}
G = \frac{\omega \mu_0 \int_V |H^2| dV}{\int_S |H^2| dS} \label{Eq_geoFactor}
\end{align}
while R\subsc{s}, R\subsc{Res} and R\subsc{BCS} account for the non-uniform field distribution. Not accounted for in the measurements are differences between the helium bath temperature and the temperature on the rf surface, which can lead to a measurable effect in Q\subsc{0} at higher field amplitudes if the bath temperature is above the $\lambda$ point of LHe at 2.17\,K \cite{Palmieri_2014}.

The high and mid-T heat treatments are processed in TRIUMF's induction furnace, which is also described in \cite{PhysRevAccelBeams.23.122001} and \cite{Dhakal:IndFurnace}. The temperature is controlled via an on/off controller set to the desired temperature. Due to the temperature measurement and regulation setup, there is an oscillation around the goal temperature as the measured temperature lags behind the temperature of the heating element. A typical mid-T bake is shown in Fig. \ref{Fig_QWR300C}, showing the heat treatment of the QWR at 300\,$^\circ$C for 3\,hours. A residual gas analyser (RGA) running during the heat treatment shows the partial pressures of selected atomic masses, which correspond to molecules and elements such as H\subsc{2}, H\subsc{2}O, CO\subsc{2}, N\subsc{2}/CO and Ar. It is important to note that the cavity ports are covered with caps made out of Nb. The purpose of these caps is to prevent line of sight from potentially contaminated furnace surfaces \cite{10.1116/6.0002624}. The caps are installed in a class 100 clean room environment after the cavity has been rinsed with high pressure pure water as would be typically done before a clean assembly. The cavity is then bagged and transported to the furnace and only just before insertion into the heating chamber the bag is removed. During the heat treatment, a number of witness samples were placed with the cavity in the furnace, but outside the rf volume of the cavity. These samples are used for analysis with secondary electron microscopy (SEM), energy-dispersive x-ray spectroscopy  (EDX), performed at the University of Victoria, and secondary ion mass spectroscopy (SIMS), measured at the University of Western Ontario. All cavity cryogenic measurements were conducted at TRIUMF's SRF facilities.
\begin{figure}[t]
	\centering
    \includegraphics[width=0.95\linewidth]{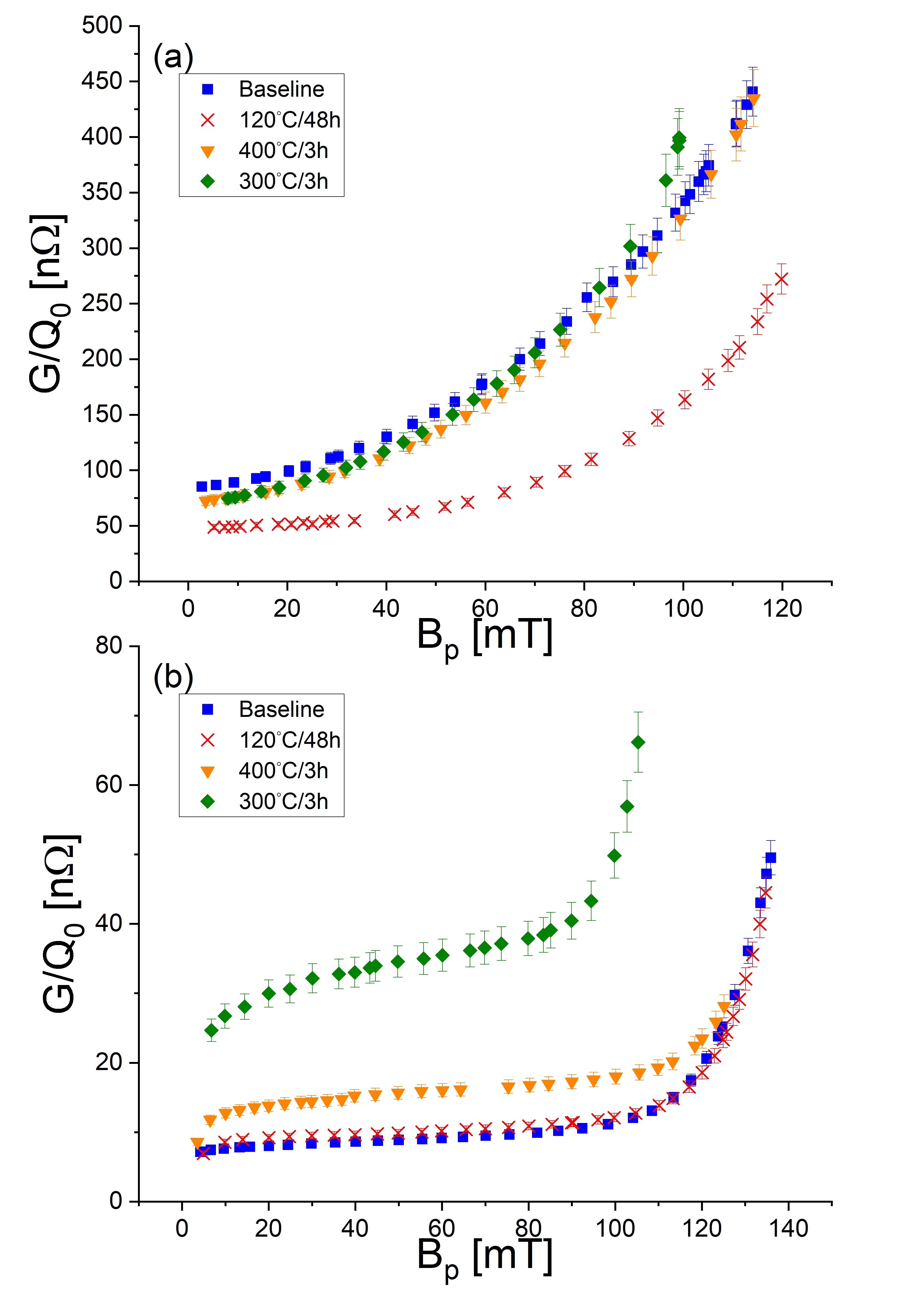}
	\caption{\label{Fig_HWR389MHz_Q_Curve} (a) 4\,K measurement of the 390\,MHz mode of the HWR show similar behaviour to the QWR fundamental mode, fig \ref{Fig_QWR217MHz_Q_Curve}, with the 120\,$^\circ$C bake out performing all others. (b) The baseline and 120\,$^\circ$C bake perform similarly, while the mid-T bakes add to R\subsc{s}.}
\end{figure}

\begin{figure}[t]
	\centering
    \includegraphics[width=0.95\linewidth]{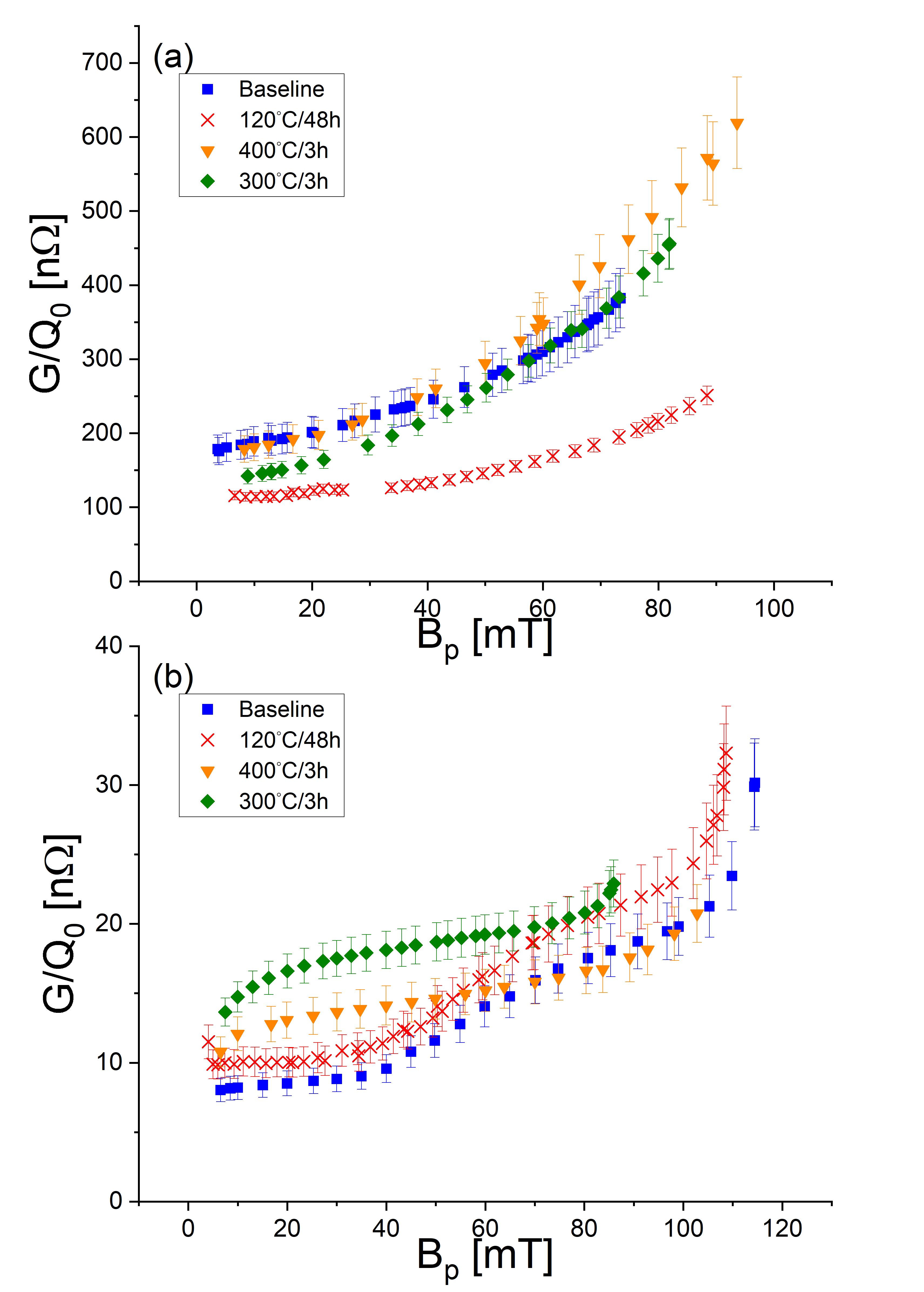}
	\caption{\label{Fig_QWR650MHz_Q_Curve} (a) At 644\,MHz and 4\,K first differences between the mid-T bakes and the baseline become noticeable, (b) but at 2\,K both the baseline and 120\,$^\circ$C still produce the highest Q\subsc{0} at low field.}
\end{figure}

The 120\,$^\circ$C bake \cite{Ciovati:120Cbake} has previously been shown to be helpful in eliminating the so-called high field Q slope, but also reduces the temperature dependent component of the surface resistance. This treatment is performed right after the baseline measurement and inside the test cryostat with resistive heaters strapped to the cavity. In this case, the heaters and temperature sensors are closer to each other, avoiding a strong temperature swing around the regulation temperature. The cavity rf space is actively pumped during the bake and throughout the cryogenic test. During the heat treatment, the surrounding helium bath space is evacuated to a low mTorr pressure to prevent heat losses and oxidation of the outer cavity surface. As the bake is done in-situ, the rf space is not exposed to air after the bake. Witness samples were processed separately in a sample vacuum oven. 

\begin{figure}[t]
	\centering
    \includegraphics[width=0.95\linewidth]{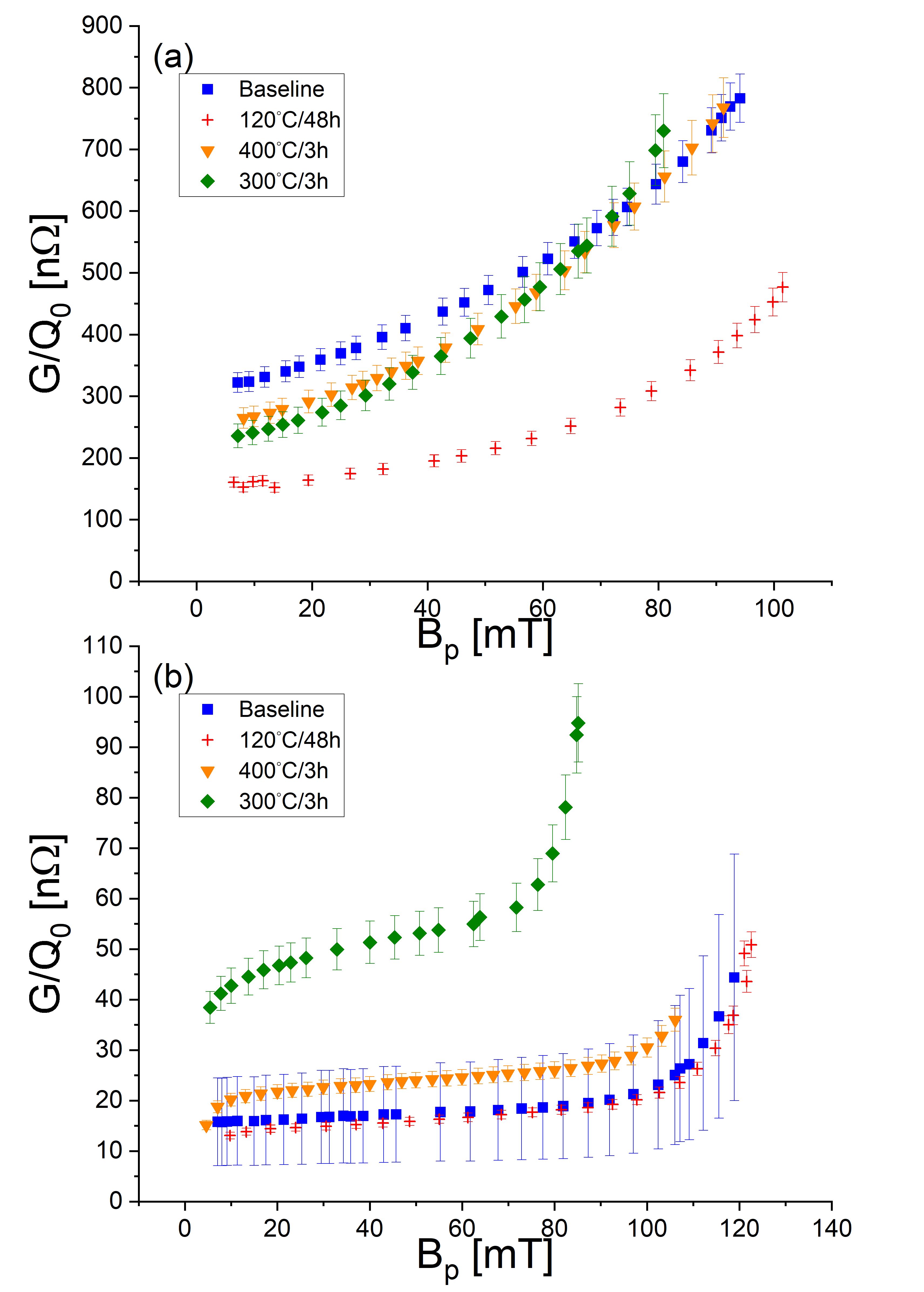}
	\caption{\label{Fig_HWR780MHz_Q_Curve} (a) At 780\,MHz and 4\,K the mid-T bakes separate further from the baseline over a larger range of rf fields. (b) At 2\,K, the baseline and 120\,$^\circ$C out perform the mid-T bakes.}
\end{figure}

\section{Results\label{secResults}}
\begin{figure}[t]
	\centering
    \includegraphics[width=0.95\linewidth]{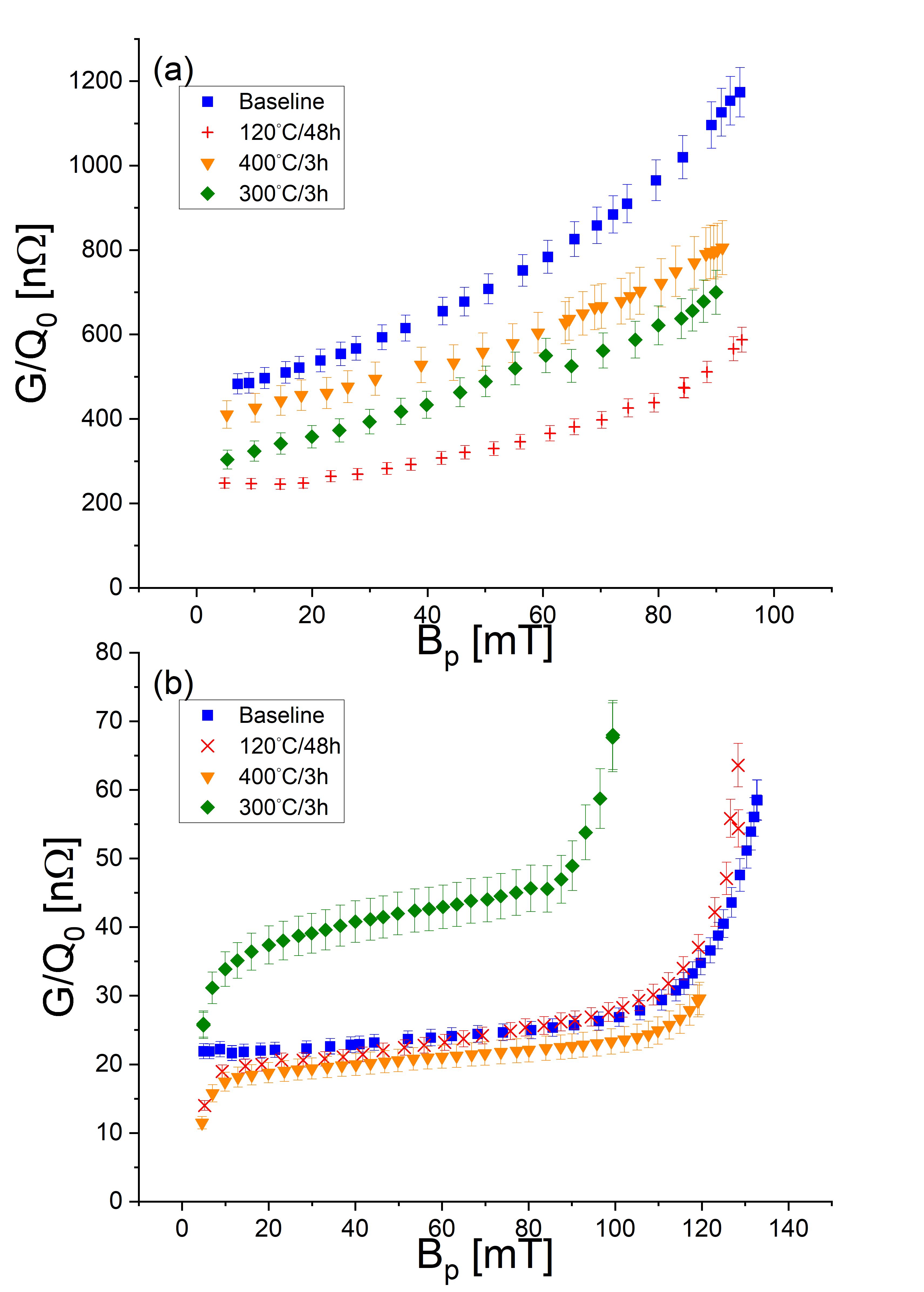}
	\caption{\label{Fig_HWR1166MHz_Q_Curve} (a) At 1166\,MHz, the mid-T bakes show a clear improvement in Q\subsc{0} compared to the baseline at 4\,K. (b) At 2\,K the 400\,$^\circ$C bake even out performs the conventional treatments.}
\end{figure}

\subsection*{Cavity Performance\label{subsecCavityPerformance}}
Cavity performance measurements were done using standard self-excited loop cw rf measurements. Both 4.2\,K and 2.0\,K measurements were performed and the results of these fixed temperature characterizations are shown in figures \ref{Fig_QWR217MHz_Q_Curve}-\ref{Fig_HWR1166MHz_Q_Curve} for all discussed treatments. It is notable that the 120\,$^\circ$C/48\,hrs bake stands out as the best performance overall at 4\,K and matches the other treatments at 2\,K. The mid-T bakes react differently to the change in rf frequency than the baseline or 120\,$^\circ$. 

At the lowest frequencies, 220 and 390\,MHz (Figs. \ref{Fig_QWR217MHz_Q_Curve} and \ref{Fig_HWR389MHz_Q_Curve}), the mid-T bakes match the baseline performance in Q\subsc{0} during the 4\,K measurements with some marginal changes in quench field. Decreasing the helium bath temperature to 2\,K, the baseline and 120\,$^\circ$C treatments show almost identical performance with G/Q\subsc{0} less than 10\,n$\Omega$ until high gradient. The two mid-T bakes perform worse both in Q\subsc{0} and quench field, with the 300\,$^\circ$C bake increasing G/Q\subsc{0} by about 20\,n$\Omega$. The decrease in Q\subsc{0} at these low temperatures and low frequencies indicates a strong increase in the temperature independent component R\subsc{Res} of the surface resistance.

In the next higher frequency modes, 644 and 780\,MHz (Figs. \ref{Fig_QWR650MHz_Q_Curve} and \ref{Fig_HWR780MHz_Q_Curve}), again the 120\,$^\circ$C bake is generally the treatment with the highest Q\subsc{0} values, especially at 4\,K. Again the mid-T bakes show a similar performance to the baseline at 4\,K, but at low field the Q\subsc{0} is increased. At 780\,MHz a larger Q\subsc{0} is measured up to a peak surface field of around 70\,mT. The lower temperature measurement again reveals a poor performance in the 300\,$^\circ$C bake, again indicating a raised R\subsc{Res}. 

At 1166\,MHz and 2\,K (Fig. \ref{Fig_HWR1166MHz_Q_Curve}), the 400\,$^\circ$C mid-T bake outperforms the other treatments. Deconvolution of R\subsc{s} (Fig. \ref{Fig_QWR_1166MHz_RTd2K}) shows that both mid-T bakes have a much reduced field dependence of R\subsc{BCS} but with an increased R\subsc{Res}. No decrease of either component with increasing rf amplitude, a signature of mid-T bakes in EP'ed 1.3\,GHz elliptical cavities, is observed.

\subsection*{Sample Measurements\label{subsecSample}}
The witness samples are 12x12\,mm\supsc{2}x1.5\,mm flat coupons made out of high (250+) RRR Nb material. Prior to the witness treatment they had received a deep etch via BCP and cleaning via an ultrasound bath. As the 120\,$^\circ$C cavity bake was done in the cryostat, the witness sample processing for this treatment was completed separately from the cavity in a small sample UHV oven. Therefore, separate samples for the QWR or HWR treatment only exist for the mid-T bake treatments. These samples were treated with the cavity together in the induction furnace, but outside the rf volume due to restricted access into that volume. The same samples are also used in the $\beta$-SRF beamline at TRIUMF \cite{10.1063/5.0137368} where the magnetic screening capabilities are being probed.

\begin{figure}[th]
	\centering
	\includegraphics[width=0.95\linewidth]{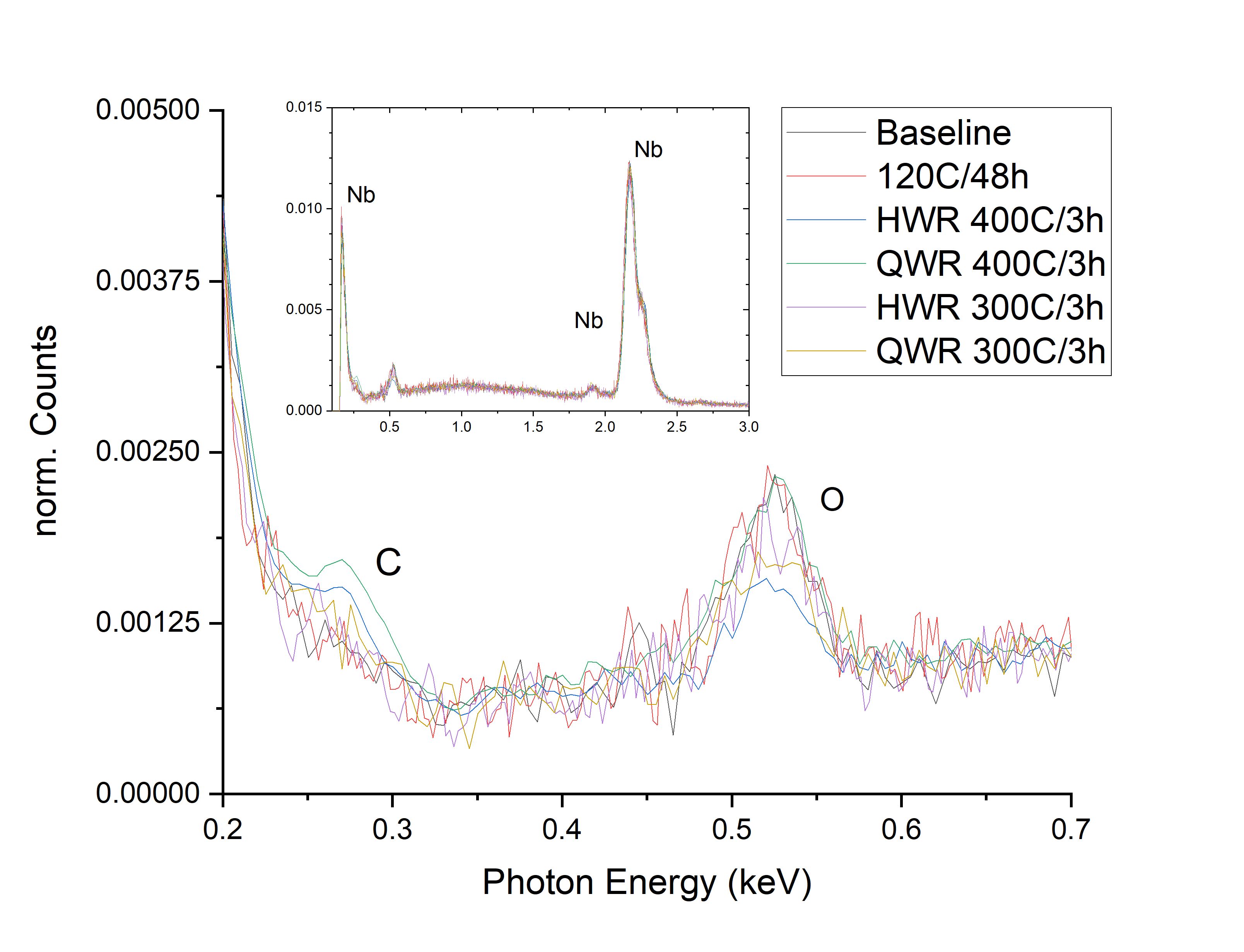}
	\caption{\label{Fig_EDX} EDX data for the baseline, 120\,$^\circ$C, and mid-T witness samples. The 400\,$^\circ$C bakes shows an enhanced carbon peak with some reduced oxygen for the HWR. The inset shows the whole energy scan with the Nb peaks.}
\end{figure}

\begin{figure}[t]
	\centering
	\includegraphics[width=0.95\linewidth]{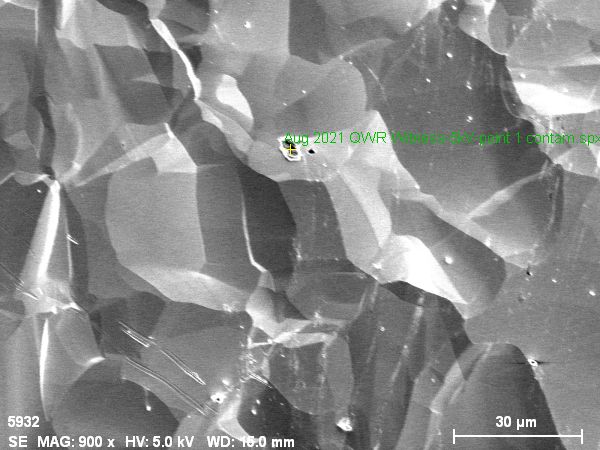}
	\caption{\label{Fig_SEMQWR} This SEM image from the QWR 300\,$^\circ$C witness sample shows the sampled area and a contaminating particulate.}
\end{figure}

Energy dispersive x-ray spectroscopy (EDX) is a technique used for an elemental characterization of the surface of samples. The sample is subjected to an electron beam. This beam can kick out electrons in the sample atoms. A higher shell electron drops down, releasing a photon of a specific energy. This allows an elemental characterization of a sample surface, given that the photon energies are known. A larger count rate indicates a larger abundance of this particular element. Fig. \ref{Fig_EDX} shows such a scan for the investigated samples. Clearly visible are the expected peaks of Nb at around 0.2, 1.9, and 2.2\,keV. The differences are in the details of the carbon and oxygen peaks at 0.25 and 0.53\,keV, respectively. In the baseline, 120\,$^\circ$C, and 300\,$^\circ$C samples any carbon signature is hidden in the tail of the Nb peak, while a noticeable increase is measurable for the two 400\,$^\circ$C samples. Oxygen is slightly suppressed in the HWR 400\,$^\circ$C and QWR 300\,$^\circ$C samples. This reduced oxygen peak is not seen in the QWR 400\,$^\circ$C sample. Overall, there seems to be only minor differences in the carbon and oxygen abundances between the baseline and mid-T bake samples. Secondary electron microscopy (SEM) of the witness samples show some contaminates, as is shown in Fig. \ref{Fig_SEMQWR} from the QWR 300\,$^\circ$C witness sample. Similar objects have been seen on the other mid-T baked samples, but not on the baseline or 120\,$^\circ$C sample.

\begin{figure}[t]
	\centering
	\includegraphics[width=0.95\linewidth]{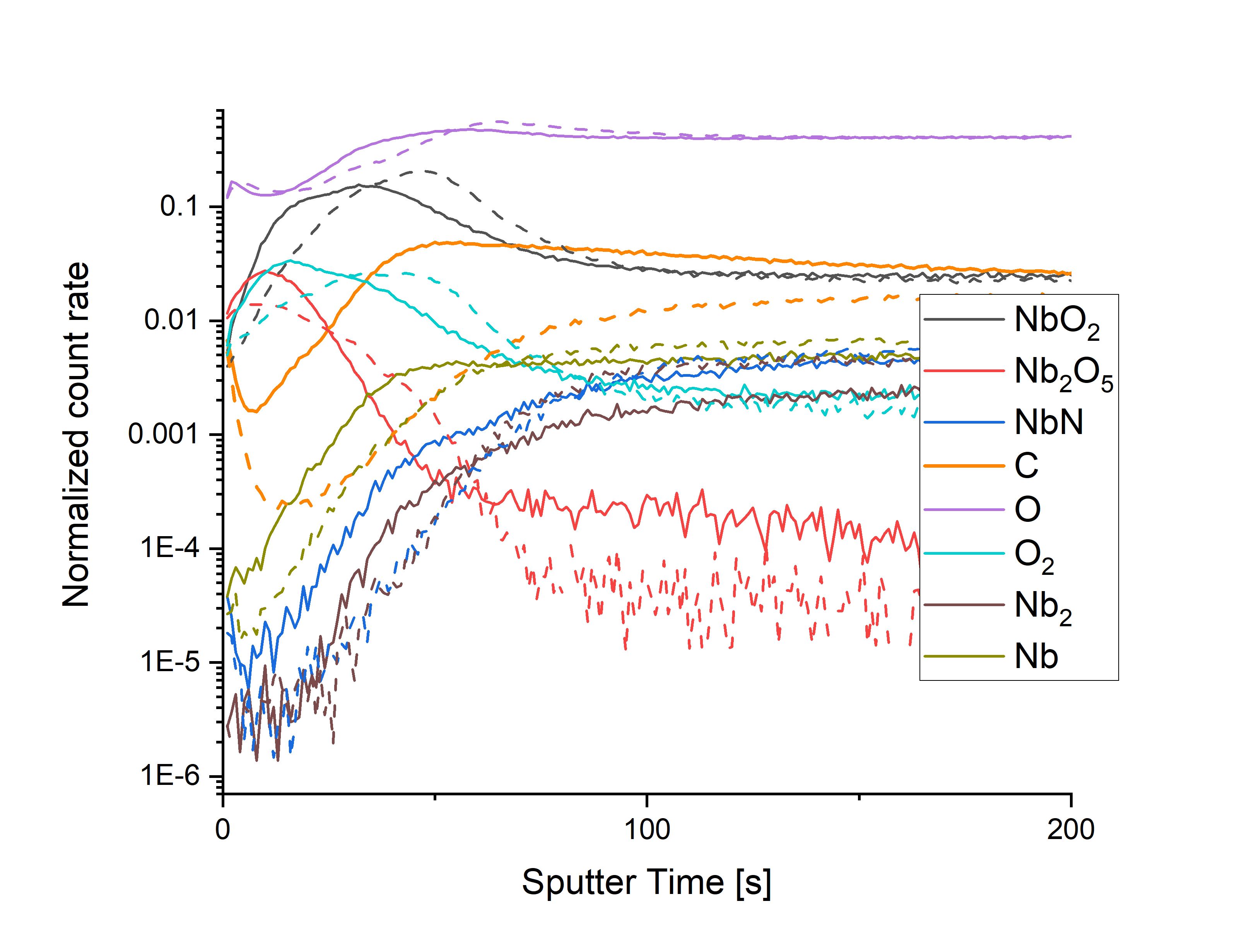}
	\caption{\label{Fig_SIMS}SIMS data of the baseline (dashed lines) and 400\,$^\circ$C (solid lines) QWR witness sample for a number of elements of interest. A notable change is in the carbon trace with a larger count rate in the mid-T sample.}
\end{figure}

The probing depth of EDX is fixed by the energy of the electron beam and depth information is not obtained. In order to get information about the elemental composition as a function depth secondary ion mass spectroscopy (SIMS) was used. In SIMS, the sample is sputtered with a heavy ion beam. For this study an ION-TOF (GmbH) TOF-SIMS IV instrument supplied a 25 keV Bi$_{3}^+$ beam pulsed at 10 kHz to bombard the sample surface in an area of about 400x400\,$\mu$m$^2$  to generate secondary ions. This releases atoms and molecules from the sample, which then can be analysed using a time-of-flight mass spectrometer. A larger sputter time results in a deeper probing of the sample. In Fig. \ref{Fig_SIMS} both the baseline and 400\,$^\circ$C QWR witness samples are analysed using this technique. A larger count rate in the carbon content is noticeable at shallower depths with the mid-T sample, while the other traces show only minor differences. The NbO\subsc{2}, Nb\subsc{2}O\subsc{5}, and O\subsc{2} signatures in the mid-T bake shift to shallower depths compared to the baseline sample. Atomic oxygen shows similar count rates between the two samples. At larger sample depths, all traces converge to similar values, showing that the heat treatment changes only the surface composition. 

A similar shift in the Nb\subsc{2}O\subsc{5} layer has been observed on samples after a mid-T bake \cite{PhysRevApplied.13.014024}. Increased carbon impurities had also been observed \cite{Yang_2023} in mid-T samples and associated with weaker rf performance.

\section{Discussion\label{secDiscussion}}

\begin{figure}[t]
	\centering
	\includegraphics[width=0.95\linewidth]{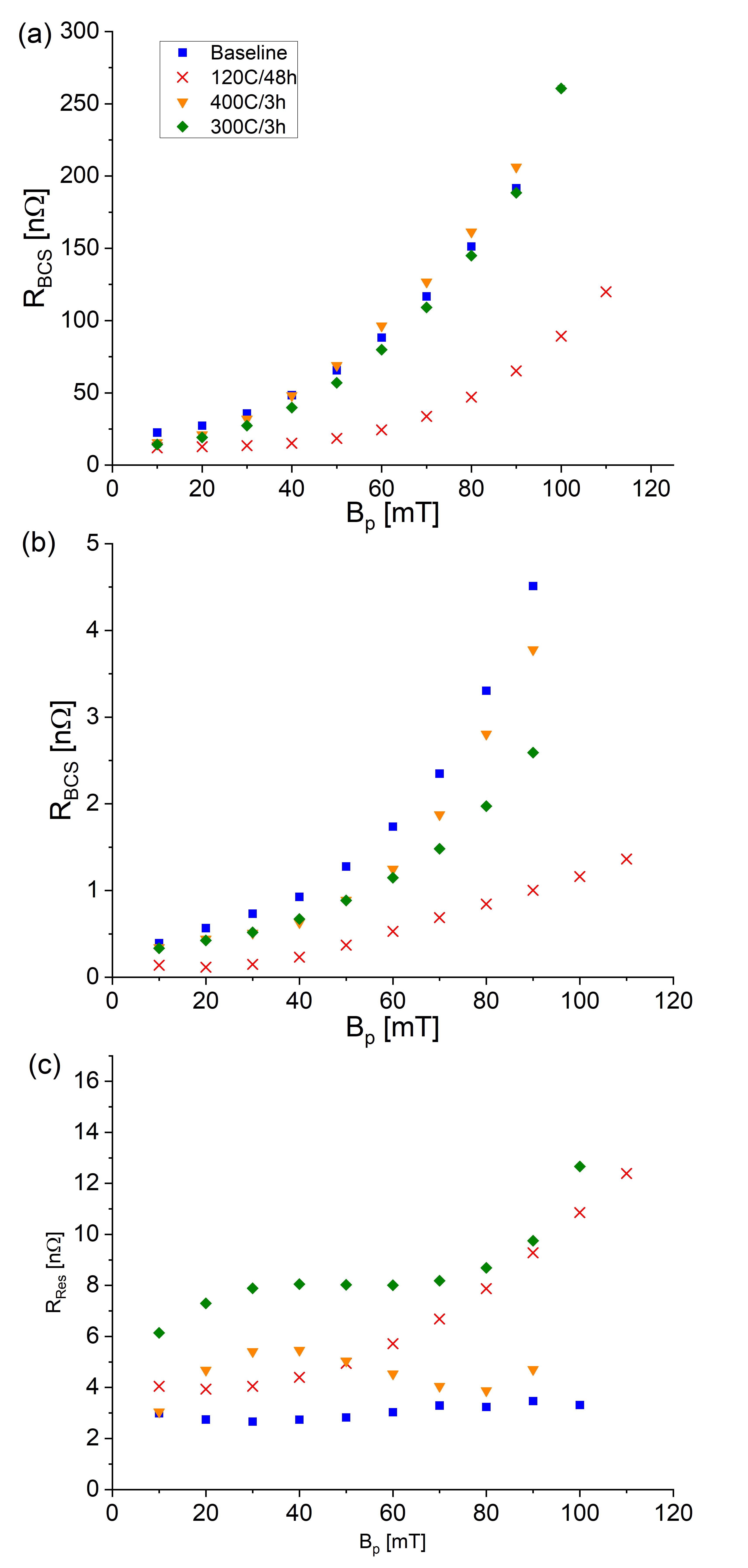}
	\caption{\label{Fig_QWR_217MHz_RTd4K}The 120\,$^\circ$C bake outperforms the other treatments at low rf frequency, 220\,MHz in this case, due to a reduced R\subsc{BCS} ((a) 4K and (b) 2K), while increasing R\subsc{Res} (c). }
\end{figure}

\begin{figure}[t]
	\centering
	\includegraphics[width=0.95\linewidth]{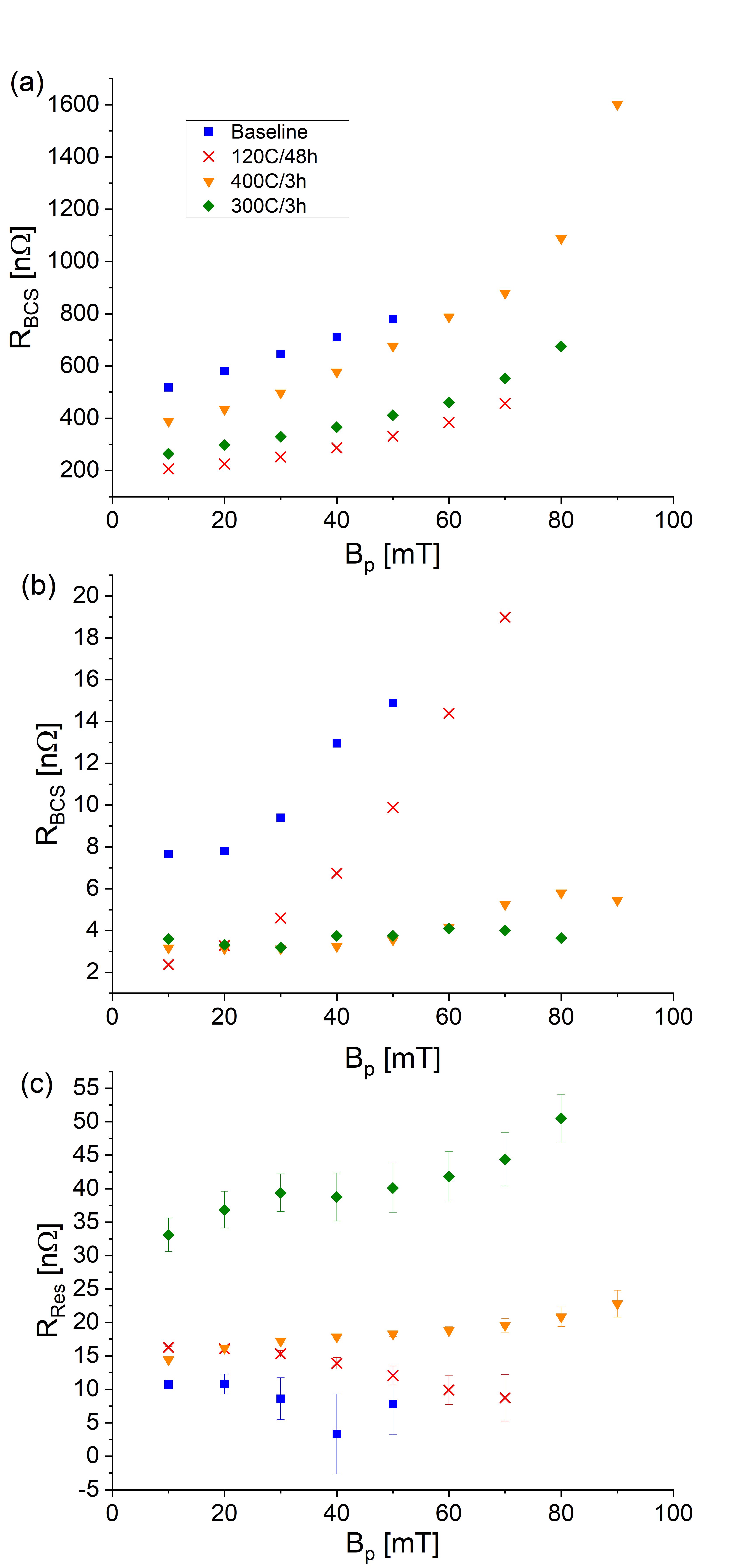}
	\caption{\label{Fig_QWR_1166MHz_RTd2K} The mid-T bakes outperform in R\subsc{BCS} the baseline and 120\,$^\circ$C bake at high rf frequency, 1166\,MHz shown here, and low temperature (b)  but a large R\subsc{Res} (c) prevents a gain in Q\subsc{0}.}
\end{figure}

The rf performance of both cavities shows clear changes between the various heat treatments, frequencies, and temperatures. To analyse the performance, the surface resistance is deconvoluted into its components R\subsc{BCS} and R\subsc{Res} by fitting R\subsc{s}(T) data collected during the cooldown from 4.2\,K to 2\,K at fixed rf peak fields B\subsc{p} from 10 up to 110\,mT to Eq. \ref{Eq_Rs}. Results of these fits are shown in figures \ref{Fig_QWR_217MHz_RTd4K} and \ref{Fig_QWR_1166MHz_RTd2K} for the 220 and 1166\,MHz modes. The baseline and 120\,$^\circ$C treatments show in all frequencies and temperatures a strong rf field dependence of R\subsc{BCS}, with the 120$^\circ$C bake producing significantly lower resistances. The residual resistance R\subsc{Res} is marginally increased as is commonly observed for this treatment. The two investigated mid-T bakes mostly follow similar trends except at low temperatures and high frequency (Fig. \ref{Fig_QWR_1166MHz_RTd2K}b). There, R\subsc{BCS} has almost no field dependence and is lower than for the 120$^\circ$C bake. A higher Q\subsc{0} compared to the 120\,$^\circ$C bake is prevented by a larger R\subsc{Res}, which could be a consequence of the higher carbon content as measured by SIMS.

\begin{figure}[tb]
	\centering
    \includegraphics[width=0.95\linewidth]{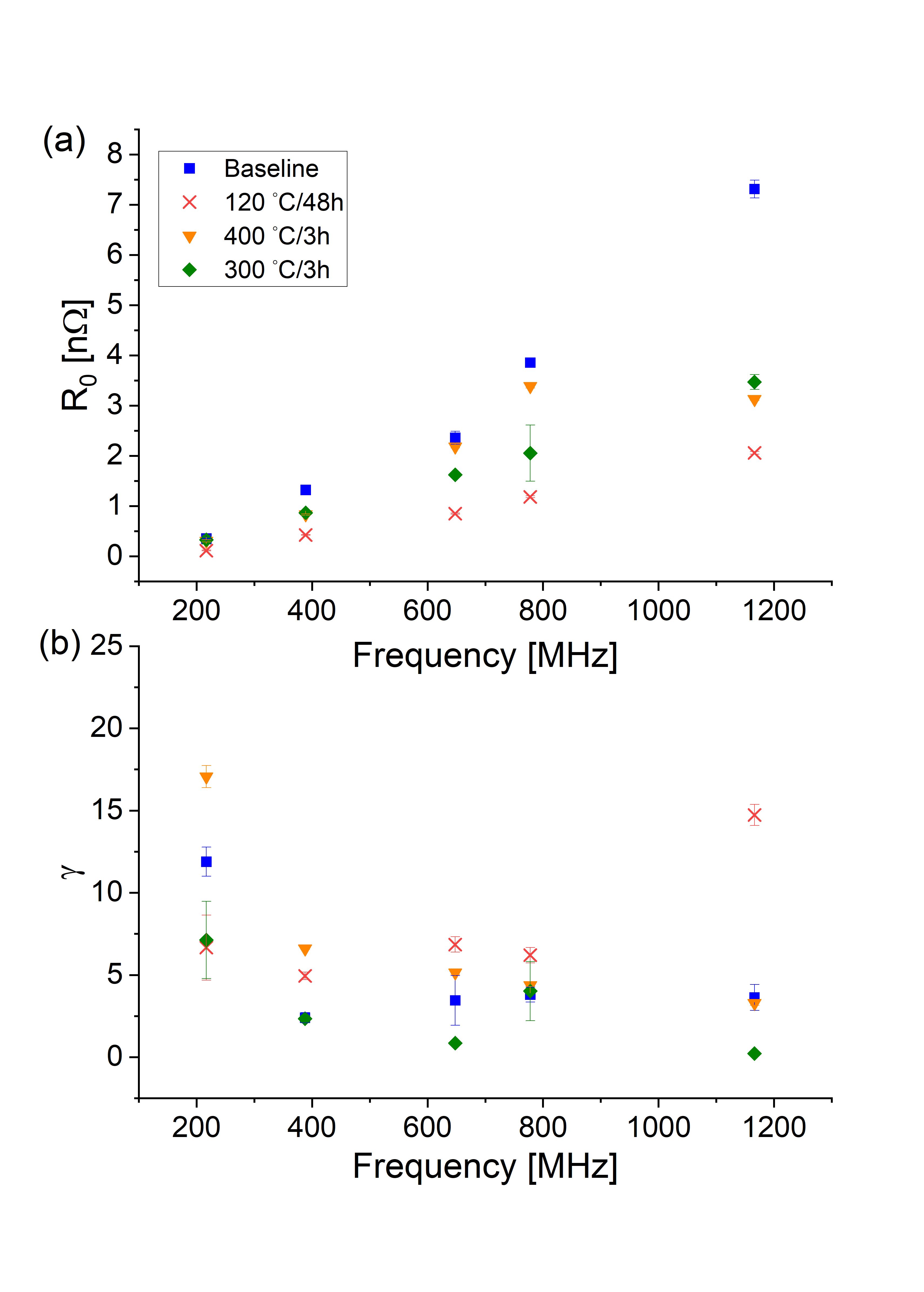}
	\caption{\label{Fig_R0_freq} (a) R\subsc{0} at 2\,K increases as the frequency increases. The 120\,$^\circ$C bake has a significantly reduced R\subsc{0} compared to the other treatments. Of note is also that the 400\,$^\circ$C bake shows a reduced R\subsc{0} at high frequency. (b) The field dependence as determined by $\gamma$ on the other hand has no clear correlation with frequency.}
\end{figure}

To quantify the dependence of R\subsc{BCS} on rf field amplitude B\subsc{p}, the data is fitted with a quadratic function according to 
\begin{align}
R_{BCS} = R_0 \left(1 + \gamma \left(\frac{B_p}{B_0}\right)^2\right) \label{Eq_gamma}
\end{align}
with $R_0$ as zero-field resistance, $B_0$ as a normalizing constant (chosen to be 100\,mT) and $\gamma$ as field dependency. The quadratic form originates from thermal feedback models \cite{GUREVICH2006_02} and other models \cite{Weingarten:PhysRevSTAB.14.101002} and is here adopted to quantify the changes in field dependence and zero-field resistance of R\subsc{BCS}. Fig. \ref{Fig_R0_freq}a shows R\subsc{0} at 2\,K resolved as function of rf frequency $f$ for the different investigated treatments. For all measured treatments and cavity modes, the baseline treatment results in the highest R\subsc{0} while the 120\,$^\circ$C bake produces the lowest zero-field resistance. The mid-T bakes fall in between. 

The baseline and 120\,$^\circ$C treatments show a steady increasing R\subsc{BCS} with frequency as would be expected from eq. \ref{Eq_Rs}, with the low temperature bake showing a lower slope, thus a reduced frequency dependence. It is noteworthy that the 400\,$^\circ$C treatment shows a lower resistance at 1166\,MHz than at 780\,MHz, which is an indication that this heat treatment is only effective at high rf frequencies above 1\,GHz. The 300\,$^\circ$C bake shows a frequency dependence of R\subsc{0} comparable to the 120\,$^\circ$C bake.

The rf amplitude dependence $\gamma$ for 2\,K shows no clear correlation with treatment or rf frequency as can be seen in Fig. \ref{Fig_R0_freq}(b). For the baseline and mid-T treatments, $\gamma$ stays roughly constant and at similar values for all but the lowest frequency mode. The 220\,MHz mode shows the highest field dependence for these three treatments. At this frequency,  $\gamma$ is significantly higher for the baseline and 400\,$^\circ$C bake compared to the other resonant modes. The 120\,$^\circ$C bake on the other hand has its highest $\gamma$ at 1166\,MHz, while $\gamma$ for the other frequencies stays constant. This together with the changes in R\subsc{0} by the treatments, shows that the differences in R\subsc{BCS} between treatments are mainly caused by the zero-field resistance R\subsc{0}.

Using all five available modes from the two cavities, the frequency dependence of R\subsc{BCS} is extracted to test against the theoretical prediction given by Eq. \ref{Eq_Rs} and codes like SRIMP which is based on \cite{HalbritterFortran}. A power law of the form 
\begin{align}
R_{BCS}(\omega) = A \omega^x \label{Eq_RTdFreq}
\end{align}
can be used to model the relationship with $A$ and $x$ as fit parameters. Based on SHRIMP, the theoretical prediction for $x$ is around 1.87. $x$ as a function of field amplitude is shown in figure \ref{Fig_RTd_freq} for 4\,K and 2\,K. The baseline treatment matches the prediction closest for a wide range of rf fields and at both temperatures. The other treatments deviate from the prediction as rf amplitude increases. At 4\,K, the baseline, 120\,$^\circ$C and 400\,$^\circ$C all follow a similar downward trend, with the 300\,$^\circ$C treatment resulting in significantly lower values than the other treatments. At 2\,K, both mid-T treatments show a reduced frequency dependent R\subsc{BCS} compared to the baseline, 120\,$^\circ$C bake or the theoretical prediction. This is an indicator of a changed layer composition in the oxygen diffused cavities, so that the resistance deviates from the typical assumed Mattis-Bardeen theory. It should be noted that the theory, and the resulting model for the surface resistance, is a low field approximation, and may not be valid at higher rf fields.

The sample analysis done with SIMS shows a modification of the oxide layer due to the mid-T bake, which is consistent with previous reports \cite{10.1063/5.0059464,Yang_2023,PhysRevApplied.13.014024}. The significant CO\subsc{2} partial pressure during the ramp up of the heat treatment (Fig. \ref{Fig_QWR300C}) and the strong carbon concentration in the SIMS measurement indicate that the cavity surfaces could have been contaminated at that time. It is speculated that this contamination causes the high R\subsc{Res} in the mid-T baked cavity measurements. 

The changed frequency response of R\subsc{BCS} with the mid-T bakes shows that the treatments have an effect on the cavity performance and that these treatments are strongly rf frequency dependent. This effect and the lack of a decreasing R\subsc{BCS} is similar to the frequency response of N-doped cavities as EP'ed 650\,MHz cavities do not show a decreasing R\subsc{BCS} \cite{10.1063/5.0068531}. A similar observation was made in a 650\,MHz EP'ed cavity with a 300\,$^\circ$C furnace bake \cite{app12020546}. Only high rf frequencies above 1\,GHz seem to be able to produce this characteristic. In addition, to get the highest Q\subsc{0} out of N-doping, EP is required over BCP \cite{PhysRevAccelBeams.22.122002}. The grain structure seems to play a role as fine grain cavities, which are most commonly used, treated with BCP did not show the otherwise typical rise in Q\subsc{0} even at 1.3\,GHz. Mid-T furnace baked cavities at 1.3\,GHz show a similar requirement \cite{Yang_2023}.

This all leads to the conclusion that high frequency above 1\,GHz and electro-polishing are required to measure decreasing surface resistances in oxygen diffused cavities. 

\section{Summary \label{secSummary}}

In this paper, the performance of two coaxial multi-mode cavities after four different treatments is presented giving a unique insight into the frequency dependence of R\subsc{BCS}. The investigated baseline and 120\,$^\circ$C bakes are established treatments, while mid-T bakes are novel and have not been reported on for low frequency coaxial type cavities before. Mid-T bake recipes have been shown to produce cavity performance similar to N\subsc{2} doped cavities. Two different mid-T treatment temperatures, 400 and 300\,$^\circ$C, were applied to the QWR and HWR multi-mode cavities and their performance measured at 4.2 and 2.0\,K. In addition, Q\subsc{0} measurements during the cooldown were done to extract the temperature dependence of the surface resistance. Measurements were done in five resonant modes, spanning from 200\,MHz to 1.2\,GHz, covering a wide range of relevant frequencies for SRF particle accelerators.

\begin{figure}[t]
	\centering
    \includegraphics[width=0.95\linewidth]{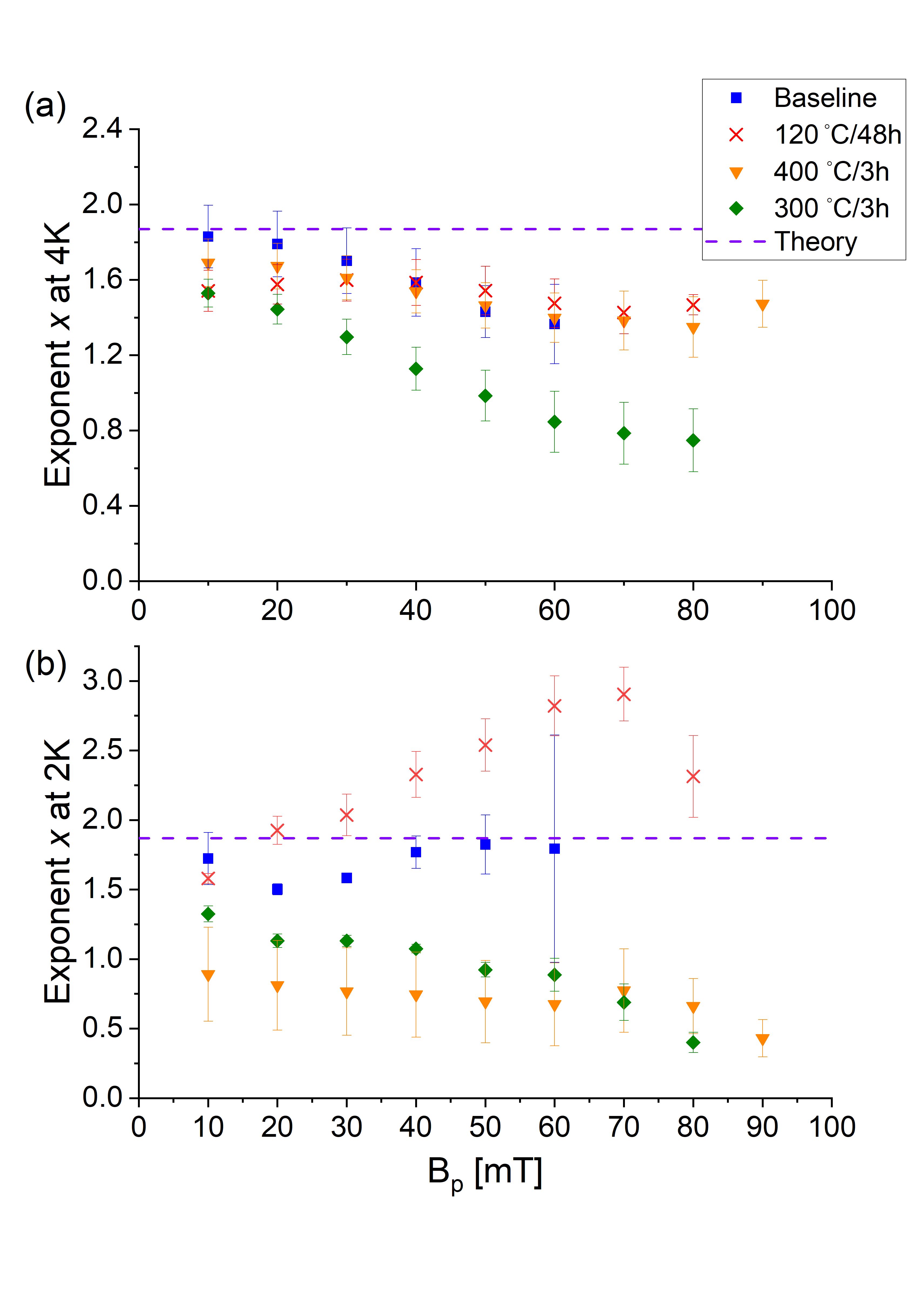}
	\caption{\label{Fig_RTd_freq} The frequency dependence exponent $x$ (eq. \ref{Eq_RTdFreq}) at (a) 4\,K and (b) 2\,K as a function of rf field amplitude. At 4\,K and higher rf fields the 300\,$^\circ$C bake shows a much lower frequency dependence than the other treatments, while at the lower temperature the trends separate between the established treatments and the newer mid-T bakes.}
\end{figure}

In the low frequency modes at 220 and 390\,MHz, the mid-T bake treatments at 400 and 300\,$^\circ$C did not improve the performance over either the baseline or the 120\,$^\circ$C bake. At 4\,K, the 120\,$^\circ$C bake produces the highest Q\subsc{0} in all cases and therefore the lowest R\subsc{BCS}. As rf frequency increases, the mid-T bakes separate from the baseline performance and perform better. At 1.2\,GHz and 2\,K, R\subsc{BCS} is lower than what was measured for the 120\,$^\circ$C bake, but no decreasing resistance with rf field could be observed as it has been in elliptical EP'ed cavities. There is a clear difference in the frequency dependence between the established treatments and the mid-T bakes, especially at higher rf fields as figure \ref{Fig_RTd_freq} shows. All this leads to the conclusion that there is a strong impact of the rf frequency on the performance of cavities with a mid-T bake.

Witness sample measurements using EDX and SIMS show some changes in the elemental composition and oxide layers. EDX shows that some of the mid-T samples have a heightened carbon content on the surface. This is confirmed with the SIMS measurement of a 400\,$^\circ$C mid-T witness sample. A source of the carbon could be the UHV furnace used for the mid-T bakes as the RGA spectrum during the heat treatment shows a significant CO\subsc{2} peak during the temperature ramp up as shown in figure \ref{Fig_QWR300C}. In addition, changes to the oxide layer were measured which are consistent with the oxide diffusion treatment. The lack of rising Q\subsc{0} in these BCP'ed cavities leads to the hypothesis that electro-polishing, and the resulting smoother surface, is needed to reach the full potential of this treatment. This might be a similar phenomenon to N\subsc{2} doping where BCP'ed cavities do not show the increase in Q\subsc{0}.

\section*{Acknowledgements}

The authors would like to thank the technical members of the SRF group, namely Devon Lang, James Keir, Bhalwinder Waraich, Ruminder Sekhon, and cryogenics group, namely David Kishi, Johnson Cheung, Rowan Bjarnson, at TRIUMF who helped tremendously with performing the cavity preparations and tests, as well as the UWO team for the SIMS measurements. Funding is provided by Natural Sciences and Engineering Research Council of Canada (NSERC) grants SAPPJ-2020-00030 (holder T. Junginger) and SAPIN-2019-00040, SAPPJ-2017-00035, and SAPEQ-2015-00005
 (holder R.E. Laxdal).
\bibliographystyle{Frontiers-Vancouver}
\bibliography{References}
\end{document}